# Comparison of solid state crystallization of boron polymorphs at ambient and high pressures.

Oleksandr O. Kurakevych[a,b]*, Yann Le Godec[a], Tahar Hammouda[c] and Céline Goujon[d]

[a]*IMPMC* & [b]*CMCP, CNRS, Université P&M Curie, F-75005 Paris, France* ; [c]*Laboratoire Magmas et Volcans, CNRS, Université Blaise Pascal, F-63000 Clermont- Ferrand, France* ; [d]*Institut Néel, CNRS, Université Joseph Fourier, F-38042 Grenoble, France*

*Corresponding author. Email: oleksandr.kurakevych@impmc.jussieu.fr

**Abstract**

Here we report the systematic study of solid-state phase transformations between boron polymorphs: α-$B_{12}$, β-$B_{106}$, γ-$B_{28}$, T-$B_{52}$ and amorphous boron (am-B). It is evident that the Ostwald rule of stages plays an important role during phase transformations not only of amorphous boron, but also of crystalline forms. We have observed the crystallization of tetragonal boron T-$B_{52}$ from amorphous phase of high purity (99.99%), which, however, cannot be easily distinguished from $B_{50}C_2$ boron compound. Many factors influence the transformations of amorphous phase, and it is possible to observe not only well known am-B → α-$B_{12}$ and am-B → β-$B_{106}$ transformations, but also am-B → T-$B_{52}$, never reported so far. At ~14 GPa the crystallization order becomes β-$B_{106}$ → α-$B_{12}$ → γ-$B_{28}$, while at ~11 GPa the intermediate crystallization of T-$B_{52}$ still was observed. This unambiguously indicates that α-$B_{12}$ is more thermodynamically stable than β-$B_{106}$ at high pressures, and renders possible to transform, at least partially, common β-phase of high purity into α-$B_{12}$ at very high pressures and moderate temperatures (below 1600 K), i.e. outside the domain of its stability.

*Keywords*: Boron; X-ray diffraction; Phase transformation

**Introduction**

High pressure (HP), although an extreme condition for a human being [1], is very common in nature, e.g., in the interiors of planets. With discovery of artificial diamond, this parameter became important not only for fundamental science, but also for industry [2,3]. It influences both static and dynamic properties of the matter. For example, the diamond structure of carbon becomes stable at high pressure, and is usually called "dense high-pressure phase". At the same time, the pressure renders carbon self-diffusion slow but activates the systematic displacements of atoms in graphite, giving rise to martensitic mechanism [4]. The impact of pressure on transport properties is a promising tool to control not only the microstructure of materials [5], but also their composition [6]. It may be explained by the fact that mainly diffusional processes lead to the phase segregation to thermodynamically stable compounds.

At the same time, displacive mechanisms may conserve the composition and metastability of the resulting phase. For example, the extreme metastability [6] of multiple graphite-like precursors of various compositions gives many opportunities to obtain multiple metastable and stable materials. It was recently illustrated by synthesis of a number of diamond-like phases of various elemental compositions, such as c-$BC_xN$ [7,8], $BC_x$ (up to $x = 5$) [9,10], etc. Most of them are metastable and nanostructured.

It was just a short description of the background of the present work, on the example of carbon-like materials, much studied under high pressure. Boron-rich solids, including the known polymorphs of boron, have also interesting behavior under extreme conditions, although not so evident to observe as in the case of carbon-like solids [11,12].

Being known for already two centuries, boron and its compounds still remain, probably, the most challenging element for fundamental and applied research in physics, chemistry and materials science [12,13]. Recently the high-pressure behavior of boron has widely attracted the attention and raised a number of high-impact results such as the high-pressure superconductivity [14], *p-T* phase diagrams [15,16], unusual high-pressure polymorphs [15,17,18] with extreme hardness [17,19], lattice behavior [20] etc. Still the knowledge of intrinsic mechanism of phase transformations between boron polymorphs remains a challenging target from both theoretical and experimental point of view, although some attempts have been already made [21-23].

α-$B_{12}$ and β-$B_{106}$ crystallize in the *R-3m* space group (trigonal syngony, $a = 4.927$ Å, $c = 12.564$ Å for α-$B_{12}$ [24] and $a = 10.947$ Å, $c = 23.903$ Å for β-$B_{106}$ [25]). The structure of α-$B_{12}$ is a distorted cubic face-center (cfc) packing of icosahedra (rhombohedral angle α ~ 58° instead of 60°, as with the ideal packing); while in the case of β-$B_{106}$, it is a packing of $B_{12}(B_5)_{12}$ groups (rhombohedral angle α ~ 65°). The principal difference between two phases consists in the presence of individual B atoms in the structure of β-$B_{106}$, which do not make a part of neither icosahedra nor more complicated clusters. The partial occupancy of the Wickoff positions of individual atoms causes the inner disorder of the lattice and can be easily recognized by weak and wide overlapped Raman bands [12] even for the samples of high crystallinity (narrow well-defined X-ray diffraction lines of ideal $d_{hkl}$ position).

Recently discovered high-pressure orthorhombic γ-$B_{28}$ phase can be also considered as a cfc packing of icosahedra with $B_2$ dumbbells in each octahedral cavity, forming in total distorted NaCl-type lattice with two differently charged cfc sublattices [15,26], giving rise to covalent polar (partially ionic) bonds between different atoms of boron. It crystallizes in the *Pnnm* space group [15] ($a$ = 5.054 Å, $b$ = 5.612 Å and $c$ = 6.966 Å) and does not contain inner disorder, as shown by sharp and well-defined Raman spectrum (Fig. 1a).

To present time, the T-$B_{52}$ (or T-$B_{50}$) phase was believed to be hypothetical boron phase, which can exists only as $B_{50}N_2$, $B_{50}C_2$ (Fig. 1b&c, open symbols) or some other compounds [12,27].

**Thermodynamic stability of boron polymorphs.**

For a long time the α-$B_{12}$ phase was believed to be thermodynamically stable at low temperatures, while disordered β-$B_{106}$ phase was considered as the high temperatures polymorph. However, detailed analysis using *ab initio* calculations render this simple interpretation somewhat ambiguous and predicts that denser α-$B_{12}$ could be stable only at high pressures [15,28]. At ambient conditions α-$B_{12}$ and β-$B_{106}$ have similar static energies, but β-$B_{106}$ has lower zero-point vibrational energy making it stable at 0 K [28], while its 0-K configurational entropy is non-zero making it stable at low and intermediate temperatures (~1.65·$10^{-4}$ eV $K^{-1}$ per atom or 15.9 J $mol^{-1}$ $K^{-1}$ [29]). However, at pressures above several GPa, much denser α-$B_{12}$ should be definitively more stable at low and even moderate temperatures [15]. Thus, with pressure increase, the free enthalpy of alpha phase is below the corresponding value for beta phase up to relatively high temperatures. This suggests the possibility of formation of alpha phase in parent beta boron bulk at high pressure, even in the *p-T* domains where they both become metastable.

At pressures higher than 10 GPa the γ-$B_{28}$ phase becomes stable up to at least 2200 K, while above this temperature T-$B_{192}$ phase has been found to be stable (see [15] and references inside). However, the high-temperature part of the diagram remains not very clear to present time.

**Crystallization of amorphous boron: Ostwald rule of stages.**

Crystallization of metastable polymorphs of boron prior to stable modifications is known for already a long time and is typical for boron. This phenomenon is justified by the Ostwald rule of stages, which insists on possibility to form and isolate the phases with intermediate values of free Gibbs energy, before the system has passed into the thermodynamically stable state with the most stable crystal structure [30,31].

Crystallization of amorphous boron in solid state by heating in an inert atmosphere (or in vacuum) has been much studied in the past [22,23,32]. It has been established that the crystallization temperature and resulting phase is a complex function not only of the short-order structure and purity of initial amorphous phase, but also of time-temperature profile (heating rate, etc.).

At ambient pressure, the crystallization of amorphous powders of pyrolytic and electrolytic samples of boron has been observed at the same temperature [32]. Pyrolytic amorphous boron crystallized as α-$B_{12}$, while electrolytic boron crystallizes directly as the thermodynamically stable β-rhombohedral modification, β-$B_{106}$ without going through the intermediate stage of α-$B_{12}$. These distinctions in the crystallization processes are indirectly brought about by the different impurity contents of the initial amorphous boron. However, the main factor that affects the crystallization processes was suggested to be the short-range order of the atom arrangement in the microstructure [32,33]. The nucleation of alpha phase is supposed to have lower activation energy, and this is one of the reason why the slow heating lead to the recovery of alpha boron after experiment. As for the crystallization of amorphous boron, the high sensitivity of the resulting phase of crystallization to contaminations has been established.

We performed the experiments with amorphous boron up to very high purity (both 99% and 99.99%). Till now only the direct crystallizations of α-$B_{12}$ and β-$B_{106}$ were observed by DTA as a function of heating rate and origin of initial amorphous phase. Our data have shown that tetragonal phase crystallizes in both cases when oxygen free hydrogenated argon was used as atmosphere. This phase has the crystal structure of T-$B_{52}$ type, never observed to present time in the case of pure boron (but mainly as $B_{50}C_2$ and $B_{50}N_2$ compounds). The lattice parameters of a synthesized phase in comparison with literature data for boron carbides $B_{50}C_x$ [34] are given in Fig. 1b&c. It is evident that the *a/c* ration is very different from those of phases obtained by CVD method. The thorough analysis of X-ray data has shown that a small amount of other phases (small amount of β-$B_{106}$ and, most probably $B_6(O/N/C)_x$ are also visible, see Fig. 2a). Their presence even in the sample obtained from so-called "99.99%- purity" boron is a striking feature. This may raise some doubt on the existence of the corresponding polymorph of boron T-$B_{52}$, and even on the possibility to use amorphous boron for the synthesis of phases pure of carbon/oxygen.

The Raman spectrum of highly crystalline (according to X-ray diffraction) T-$B_{52}$ phase (bottom spectrum of Fig. 1a) show a superposition of wide weak bands and seems to correspond to the structure with inner disorder of atoms allover some Wyckoff positions (like in the case of β-$B_{106}$ [25] and $B_{50}X_2$ phases [27]).

Our results presented in this chapter give a strong support to the previous reports on the strong sensitivity of crystallization product to the purity and structure of initial am-B.

**Transformations of beta phase.**

The high-purity β-$B_{106}$ melts at pressures at least up to 8 GPa [35,36], and completely transforms into γ-$B_{28}$ phase at pressures above 10 GPa and temperatures above 1800 K [15]. However, our experimental study of the beginning of the β → γ phase transformation (~1400 K) shows that the transformation does not pass directly, and the T-$B_{52}$ or α-$B_{12}$ phase crystallize as intermediate ones. For these experiments we have used the boron nitride capsules, proved to be inert in relation to crystalline boron under employed *p-T* conditions and refractory to avoid the formation of liquid phase in the system at employed temperature [37-40].

Fig. 3a shows the X-ray diffraction patterns of the samples recovered from 1400 K (14.4 GPa, 10-min heating). One can see always three phases: initial β-$B_{106}$ (partially recrystallized), intermediate α-$B_{12}$ and final γ-$B_{28}$. The Raman data (Fig. 3b) confirm the presence of three mentioned phases. At higher temperatures, α-phase completely transforms into γ-phase. Obviously, the subsequent crystallization of phases in the framework of Ostwald rule of stages occurs in the case of crystalline boron, as well as in the case of amorphous one (Fig. 4a&b). The experiment performed at lower pressure (11 GPa, 10-min heating) has shown the crystallization of T-$B_{52}$ phase as intermediate one (between β-$B_{106}$ and γ-$B_{28}$) in the beginning of phase transformation (Figs. 2b&4a). Although the results at ambient pressure raised some doubt on the existence of T-$B_{52}$, the X-ray diffraction data from the high-pressure sample (Fig. 1b&c) give a strong support to its existence. Anyway, this phase still remains the object of study. The weak and broad Raman data of this phase (bottom spectrum of Fig. 1a) is indicative of an inner disorder of boron atoms allover Wyckoff positions in the structure, similar to β-$B_{106}$ and in contrast with γ-$B_{28}$ and α-$B_{12}$. The difference in such occupancies may be the reason of the difference of lattice parameters of T-$B_{52}$ obtained at different pressures.

All kinetic results obtained in the present work (Fig. 4a) indicate that curves "chemical potential vs pressure" at the temperatures in the vicinity of the onset of transformations for boron polymorphs (~1500 K) should have mutual position as presented on Fig. 4b. This allows us to reasonably explain the observed crystallization paths in the framework of the Ostwald rule of stages. The relative positions of α-, β- and γ- phases are very similar to those predicted by ab initio calculations [15]. However, the situation with T-$B_{52}$ phase seem to be somewhat strange. According to our experiments on metastable crystallization, it should be very close to alpha phase (Fig. 4b), while *ab initio* calculations show that at 0 K the free enthalpy of T-$B_{52}$ should be much higher [15]. This discrepancy could be explained by the non-negligible 0-K configurational entropy of this tetragonal phase (due to the partial occupancy of Wyckoff positions [12,27]), which strongly decreases the free enthalpy at high temperatures, similarly to β-phase.

**Structural features of transformation mechanism.**

To present time only the mechanism of direct phase transformation of α-$B_{12}$ into β-$B_{106}$ has been studied [21]. It has been proven that the transformation is not exclusively limited by solid diffusion, but the displacive stages accompanied by the strain/stress and stacking faults accumulation along some crystallographic directions play important role. Our observations (Tab. 1) indicate that the residual strains in the recovered samples containing α-$B_{12}$, β-$B_{106}$ and γ-$B_{28}$ phases are present, although remarkably less pronounced.

When the crystal accumulates the stacking faults of the ($h_0k_0l_0$) layer (the latter remains practically unperturbed) in the direction $\mathbf{g}_{h0k0l0}$ ($\mathbf{g}_{h0k0l0} \perp (h_0k_0l_0)$), the profiles, *d*-positions and intensities of *hkl* reflections may change in different ways, e.g. positive or negative deviation of *d*-spacings, asymmetry, widening and/or weakening of certain reflections, etc. (see, for example, ref. [41] and references inside).

The lack (or very weak intensity), asymmetry and shift of certain reflections of boron phases is typical for powder diffraction patterns of our recovered samples (Fig. 4a, Tab. 1). The accumulation of stacking faults in certain crystallographic directions has been previously attributed to specific displacive feature during the alpha to beta transformation [21]. Evidently, the reverse is true (see Tab. 1), indicative that inverse transformation (β-$B_{106}$ → α-$B_{12}$) has similar features. However, the detailed study of such particularities is beyond the scope of this report and will be published elsewhere.

**Conclusions**

We performed the systematic study of solid-state phase transformations of boron polymorphs: α-$B_{12}$, β-$B_{106}$, γ-$B_{28}$ and amorphous boron (am-B). It is evident that the Ostwald rule of stages plays an important role during phase transformations not only of amorphous boron, but also of crystalline boron. We have first observed the crystallization of tetragonal boron T-$B_{52}$ from amorphous phase of high purity at ambient pressure and temperatures ~1400 K. However, no data allow distinguishing it from the $B_{50}C_2$ boron compound. Many factors influence the transformations of amorphous phase, and it is possible to observe not only well known am-B → α-$B_{12}$ and am-B → β-$B_{106}$ transformations, but also am-B → T-$B_{52}$, never observed so far. While at ambient pressure the crystallization order is am-B → (α-$B_{12}$ or T-$B_{52}$ →) β-$B_{106}$, at high pressure it becomes β-$B_{106}$ → T-$B_{52}$ → γ-$B_{28}$ (~11 GPa) or β-$B_{106}$ → α-$B_{12}$ → γ-$B_{28}$ (~14 GPa). This unambiguously indicates that α-$B_{12}$ is more thermodynamically stable than β-$B_{106}$ at high pressures, and renders possible to transform common β-phase of high purity into α-$B_{12}$ (or T-$B_{52}$) at high pressures and moderate temperatures (~1400 K).

**Experimental methods.**

Amorphous boron up to very high purity (both 99% and 99.99%, Alfa) and well-crystallized β-$B_{106}$ (99.999% purity, Alfa) were used as starting materials.

To study the phase transitions of amorphous boron at ambient pressure, differential thermal analyses (DTA) were performed by using a Netzsch Tasc 414/3 analyser. The sample temperature was increased up to 1570 K at a heating rate of 2 K/min in hydrogenated argon atmosphere, then cooled as fast as possible (60 K/min).

High pressure experiments were performed in the octahedral multi-anvil apparatus (LMV) using a Walker module [44] and following the procedure described in ref. [45]. The starting material was contained in a boron nitride container, which was subsequently inserted in a $LaCrO_3$ cylindrical furnace before being introduced into the high-pressure cell consisting of a Cr-doped MgO octahedron of 14 mm edge length. The pressure medium was squeezed by eight cubic tungsten carbide anvils with 8 mm length truncation. For each run the pressure was first raised to desired pressure in 3 hours. The temperature was subsequently raised. Dwell temperature was reached in 8 to 9 minutes and the sample was kept at high temperature for 10 minutes. Temperature was monitored using a W-Re (5/26) thermocouple. Quenching was performed by shutting off the electrical power, resulting in quenching rate of several hundreds of degrees per second. The sample was then slowly decompressed, in 12 hours.

The recovered samples were studied by X-ray diffraction. Two X'Pert PRO PANalytical powder X-ray diffractometers (Bragg-Brentano geometry) employing CoKα1

CuKα1 radiation were used. The goniometer was aligned using high purity silicon ($a = 5.431066$ Å). Unit cell parameters were derived from the LeBail profile analysis performed using the PowderCell program.

The homogeneity of the samples was established by micro-Raman spectroscopy (5 micron beam). Raman spectra were collected in the backscattering geometry using peltier-chilled Raman spectrometer JY HR800 (the 488-nm excitation laser beam). The spectrometer was calibrated using the $\varGamma_{25}$ phonon of Si (*Fd-3m*).

**Acknowledgements**

The authors thank Profs. V. L. Solozhenko, A. R. Oganov and L. Dubrovinsky for valuable discussions. The multianvil apparatus of Laboratoire Magmas et Volcans is financially supported by the Centre National de la Recherche Scientifique (Instrument National de l'INSU).

**References**

[1] P. W. Bridgman, Collected Experimental Papers, eds., Harvard University Press, Cambridge, Massachusetts, 1964.
[2] F. P. Bundy, H. T. Hall, H. M. Strong and R. H. Wentorf, *Man made diamonds.*, Nature 176 (1955), pp. 51-55.
[3] H. Liander, *Artificial Diamonds*, ASEA Journal 28 (1955), pp. 97-98.
[4] V. F. Britun and A. V. Kurdyumov, *Mechanisms of martensitic transformations in boron nitride and conditions of their development*, High Press. Res. 17 (2000), pp. 101-111.
[5] N. Dubrovinskaia, V. L. Solozhenko, N. Miyajima, V. Dmitriev, O. O. Kurakevych and L. Dubrovinsky, *Superhard nanocomposite of dense polymorphs of boron nitride: Noncarbon material has reached diamond hardness*, Appl. Phys. Lett. 90 (2007), pp. 101912.
[6] P. F. McMillan, *New materials from high-pressure experiments*, Nature Mater. 1 (2002), pp. 19-25.
[7] V. L. Solozhenko, D. Andrault, G. Fiquet, M. Mezouar and D. C. Rubie, *Synthesis of superhard cubic BC$_2$N*, Appl. Phys. Lett. 78 (2001), pp. 1385-1387.
[8] V. L. Solozhenko, *High-pressure synthesis of novel superhard phases in the B-C-N system: recent achievements*, High Press. Res. 29 (2009), pp. 612-617.
[9] V. L. Solozhenko, O. O. Kurakevych, D. Andrault, Y. Le Godec and M. Mezouar, *Ultimate metastable solubility of boron in diamond: Synthesis of superhard diamond-like BC$_5$*, Phys. Rev. Lett. 102 (2009), pp. 015506.
[10] V. L. Solozhenko, O. O. Kurakevych, D. Andrault, Y. L. Godec and M. Mezouar, *Erratum: Ultimate Metastable Solubility of Boron in Diamond: Synthesis of Superhard Diamondlike BC$_5$ [Phys. Rev. Lett. 102, 015506 (2009)]*, Phys. Rev. Lett. 102 (2009), pp. 179901.
[11] O. O. Kurakevych and V. L. Solozhenko, *High-pressure route to superhard boron-rich solids*, High Pressure Research 31 (2011), pp. 48-52.
[12] O. O. Kurakevych, *Superhard phases of simple substances and binary compounds of the B-C-N-O system: from diamond to the latest results (a Review)* J. Superhard Mater. 31 (2009), pp. 139-157.

[13]  A. R. Oganov and V. L. Solozhenko, *Boron: a hunt for superhard polymorphs*, J. Superhard Mater. 31 (2009), pp. 285-291.
[14]  M. I. Eremets, V. W. Struzhkin, H. K. Mao and R. J. Hemley, *Superconductivity in boron*, Science 293 (2001), pp. 272-274.
[15]  A. R. Oganov, J. Chen, C. Gatti, Y. Ma, Y. Ma, C. W. Glass, Z. Liu, T. Yu, O. O. Kurakevych and V. L. Solozhenko, *Ionic high-pressure form of elemental boron.*, Nature 457 (2009), pp. 863-867.
[16]  K. Shirai, A. Masago and H. Katayama-Yoshida, *High-pressure properties and phase diagram of boron*, Phys. Stat. Solidi B 244 (2007), pp. 303-308.
[17]  V. L. Solozhenko, O. O. Kurakevych and A. R. Oganov, *On the hardness of a new boron phase, orthorhombic γ-$B_{28}$*, J. Superhard Mater. 30 (2008), pp. 428-429.
[18]  R. H. Wentorf Jr., *Boron: Another Form*, Science 147 (1965), pp. 49-50.
[19]  V. A. Mukhanov, O. O. Kurakevych and V. L. Solozhenko, *Thermodynamic model of hardness: Particular case of boron-rich solids*, J. Superhard Mater. 32 (2010), pp. 167-176.
[20]  Y. Le Godec, O. O. Kurakevych, P. Munsch, G. Garbarino and V. L. Solozhenko, *Equation of state of orthorhombic boron, γ-$B_{28}$*, Solid State Comm. 149 (2009), pp. 1356-1358
[21]  P. Runow, *Study of the α to β Transformation in Boron*, J. Mater. Sci. 7 (1972), pp. 499-511.
[22]  N. E. Solov'ev, V. S. Makarov and Y. A. Ugai, *The crystallization kinetics of amorphous boron*, J. Less Comm. Met. 117 (1986), pp. 21-27.
[23]  N. E. Solov'ev, V. S. Makarov, L. N. Meschaninova and Y. A. Ugai. in Boronrich solids 532-537 (AIP, Albuquerque, NM (USA), 1991).
[24]  B. F. Decker and J. S. Kasper, *The crystal structure of a simple rhombohedral form boron*, Acta Crystallogr. 12 (1959), pp. 503-506.
[25]  J. L. Hoard, D. B. Sullenger, C. H. L. Kennard and R. E. Hughes, *The structure of beta-rhombohedral boron*, J. Solid State Chem. 1 (1970), pp. 268-277.
[26]  A. R. Oganov, J. Chen, C. Gatti, Y. Ma, Y. Ma, C. W. Glass, Z. Liu, T. Yu, O. O. Kurakevych and V. L. Solozhenko, *Addendum: Ionic high-pressure form of elemental boron*, Nature 460 (2009), pp. 292-292.
[27]  G. Will and K. H. Kossobutzki, *X-ray diffraction analysis of $B_{50}C_2$ and $B_{50}N2$ crystallizing in the "tetragonal" boron lattice*, J. Less Comm. Met. 47 (1976), pp. 33-38.
[28]  M. J. van Setten, M. A. Uijttewaal, G. A. de Wijs and R. A. de Groot, *Thermodynamic stability of boron: The role of defects and zero point motion.*, J. Am. Chem. Soc. 129 (2007), pp. 2458-2465.
[29]  A. Masago, K. Shirai and H. Katayama-Yoshida, *Crystal stability of α- and β-boron*, Phys. Rev. B 73 (2006), pp. 104102 1-10.
[30]  W. Ostwald, *Studien uber die Bildung und Umwandlung fester Korper*, Z. Phys. Chem. 22 (1897), pp. 289.
[31]  T. Threlfall, *Structural and Thermodynamic Explanations of Ostwald's Rule*, Org. Proc. Res. & Dev. 7 (2003), pp. 1017-1027.
[32]  S. O. Shalamberidze, G. I. Kalandadze, D. E. Khulelidze and B. D. Tsurtsumia, *Production of α-Rhombohedral Boron by Amorphous Boron Crystallization*, J. Solid State Chem. 154 (2000), pp. 199-203.
[33]  M. Kobayashi, I. Higashi and M. Takami, *Fundamental Structure of Amorphous Boron*, J. Solid State Chem. 133 (1997), pp. 211-214.

[34] U. Jansson, J. O. Carlsson, B. Stridh, S. Söderberg and M. Olsson, *Chemical vapour deposition of boron carbides I: Phase and chemical composition*, Thin Solid Films 172 (1989), pp. 81-93.

[35] V. V. Brazhkin, T. Taniguichi, M. Akaishi and S. V. Popova, *Fabrication of β-boron by chemical-reaction and melt-quenching methods at high pressures*, J. Mater. Res. 19 (2004), pp. 1643-1648.

[36] V. L. Solozhenko, O. O. Kurakevych, V. Z. Turkevich and D. V. Turkevich, *Phase Diagram of the B-$B_2O_3$ System at 5 GPa: Experimental and Theoretical Studies*, J. Phys. Chem. B 112 (2008), pp. 6683-6687.

[37] V. L. Solozhenko, Y. Le Godec and O. O. Kurakevych, *Solid-state synthesis of boron subnitride, $B_6N$: myth or reality?*, Comptes Rendus Chimie 9 (2006), pp. 1472-1475.

[38] V. L. Solozhenko and O. O. Kurakevych, *Chemical interaction in the B-BN system at high pressures and temperatures. Synthesis of novel boron subnitrides*, J. Solid State Chem. 182 (2009), pp. 1359-1364.

[39] V. L. Solozhenko, O. O. Kurakevych, V. Z. Turkevich and D. V. Turkevich, *On the Problem of the Phase Relations in the B–BN System at High Pressures and Temperatures*, J. Superhard Mater. 31 (2009), pp. 1-6.

[40] V. L. Solozhenko, O. O. Kurakevych, V. Z. Turkevich and D. V. Turkevich, *Phase Diagram of the B-BN System at 5 GPa*, J. Phys. Chem. B 114 (2010), pp. 5819-5822.

[41] L. Balogh, G. Ribárik and T. Ungár, *Stacking faults and twin boundaries in fcc crystals determined by x-ray diffraction profile analysis*, J. Appl. Phys. 100 (2006), pp. 023512

[42] F. H. Horn, *On the Crystallization of Simple Rhombohedral Boron from Platinum*, J. Electrochem. Soc. 106 (1959), pp. 905-906.

[43] G. Parakhonskiy, N. Dubrovinskaia, L. Dubrovinsky, S. Mondal and S. van Smaalen, *High pressure synthesis of single crystals of α-boron*, J. Cryst.Growth 321 (2011), pp. 162-166.

[44] D. Walker, M. A. Carpenter and C. M. Hitch, *Some simplifications to multianvil devices for high pressure experiments*, Amer. Mineral. 75 (1990), pp. 1020-1028.

[45] T. Hammouda, *High pressure melting of carbonated eclogite and experimental constraints on carbon recycling and storage in the mantle*, Earth Planet Sci. Lett. 214 (2003), pp. 357-368.

Tab. 1 X-ray diffraction data for the sample recovered from 14.4 GPa and 1400 K (the top powder diffraction pattern of Fig. 3a)

| d-Spacings of hkl reflections | | | | | | | | | | | |
|---|---|---|---|---|---|---|---|---|---|---|---|
| γ-B | | | | α-B | | | | β-B | | | |
| hkl | $d_{theor}$ | $d_{exp}$ | $\varepsilon_d$, % | hkl | $d_{theor}$ | $d_{exp}$ | $\varepsilon_d$, % | hkl | $d_{theor}$ | $d_{exp}$ | $\varepsilon_d$, % |
| 011 | 4.370 | 4.361 | -0.21 | 111 | 4.188 | 4.177 | -0.26 | 111 | 7.968 | 7.898 | -0.88 |
| 101 | 4.091 | 4.078 | -0.32 | 100 | 4.040 | 4.014 | -0.64 | 110 | 7.427 | 7.419 | -0.11 |
| 110 | 3.756 | 3.745 | -0.29 | | | | | 1-10 | 5.473 | 5.488 | 0.27 |
| 002 | 3.483 | 3.478 | -0.14 | | | | | 1-11 | 4.649 | 4.666 | 0.37 |
| 111 | 3.306 | 3.298 | -0.24 | | | | | 210 | 4.511 | 4.504 | -0.16 |
| | | | | | | | | 222 | 3.984 | 3.959 | -0.63 |
| Lattice parameters | | | | | | | | | | | |
| γ-B | | | | α-B | | | | β-B | | | |
| a | 5.054 | 5.035 | -0.38 | a | 5.0627 | 5.0426 | -0.40 | a | 10.17 | 10.1379 | -0.32 |
| b | 5.612 | 5.605 | -0.13 | α | 58.2339 | 58.0465 | -0.32 | α | 65.1196 | 65.6435 | 0.80 |
| c | 6.966 | 6.96 | -0.09 | | | | | | | | |

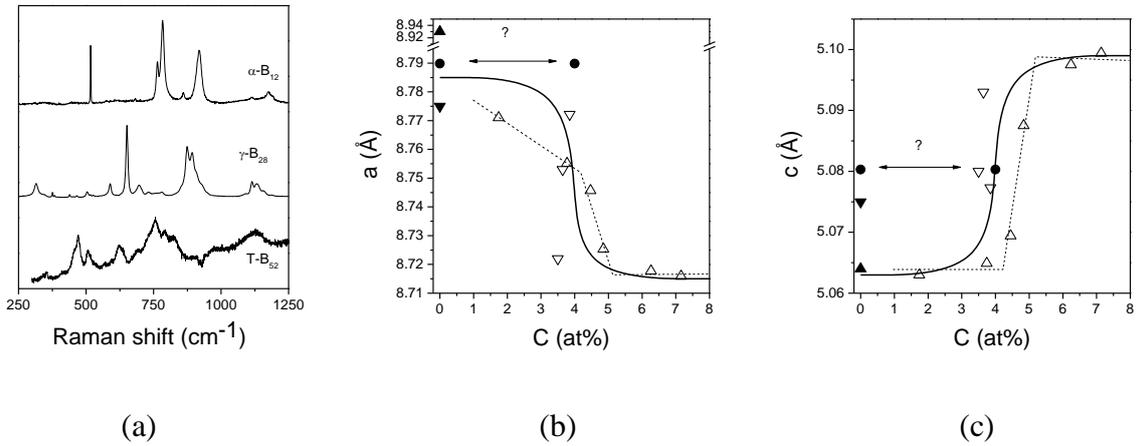

Fig. 1 (**a**) Raman spectra of α-, γ- and T-borons. (**b&c**) Lattice parameters of $B_{50}C_x$ phases as a function of composition. The dotted line with symbols Δ represent the data of [34]. Symbols ∇ represent the data of [27] (see also references inside). Solid line is a general guide for eyes. Symbols show our experimental data (● and ▼ – crystallization at ambient pressure, ▲ – crystallization at high pressure).

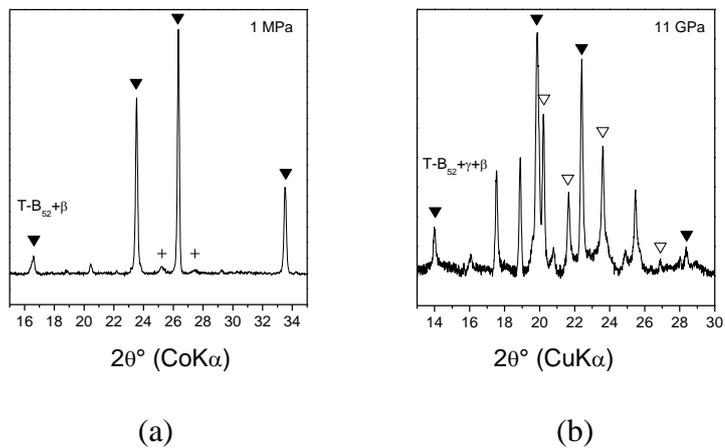

Fig. 2. X-ray powder diffraction pattern of samples recovered after DTA experiment at ambient pressure (a) and from 11 GPa and 1500 K (b). The symbols are: ∇ - γ-$B_{28}$, ▼ - T-$B_{52}$. + - reflections of $B_6$(O/N/C) (?). All non-marked reflections correspond to β-$B_{106}$.

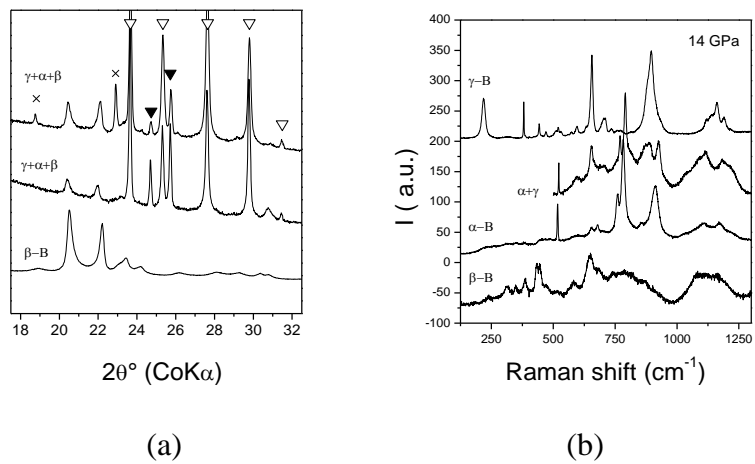

Fig. 3. (**a**) X-ray powder diffraction patterns of starting β-boron sample (bottom) and of samples recovered from 14.4 GPa and 1400 K. The symbols are: × - reflections of β-$B_{106}$ that became visible, ∇ - γ-$B_{28}$, ▼ - α-$B_{12}$. (**b**) Raman spectra collected in different places of β-$B_{106}$ recovered from 14.4 GPa and 1400 K.

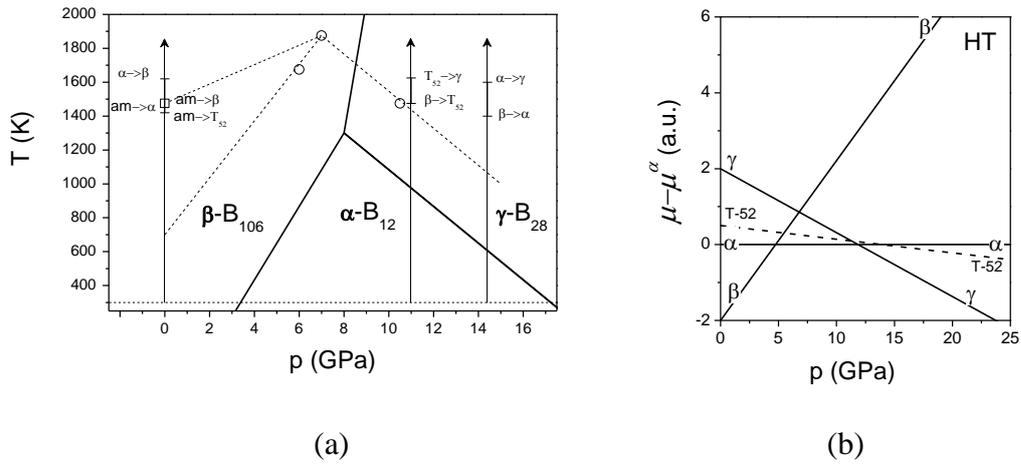

Fig. 4. (**a**) Crystallization order of boron forms at ambient and high pressure. Solid lines represent the phase diagram of boron obtained by combination of *ab initio* calculations and experimental data [15]. The dashed lines show the data on the α-boron crystallization (the highest temperatures) from liquid solutions (□ from ref. [42], ○ from ref. [43]). (**b**) Schematic representation of the relative positions of chemical potentials of α-, β-, γ- and T-52 polymorphs of boron at temperatures above 1000 K.